\documentclass[twoside,11pt]{article}

\usepackage{amsmath}
\usepackage{algorithm}
\usepackage[noend]{algpseudocode}
\renewcommand{\algorithmicrequire}{\textbf{Input:}}
\renewcommand{\algorithmicensure}{\textbf{Output:}}

\usepackage{booktabs}
\usepackage{tabularx}
\usepackage{caption}
\usepackage{subcaption}
\usepackage{color}
\usepackage{macros}
\usepackage{relsize}
\usepackage{placeins}

\usepackage{jmlr2e}
\usepackage[colorlinks=false,allbordercolors={1 1 1}]{hyperref}

\jmlrheading{15}{2014}{2087-2112}{10/12; Revised 1/14}{6/14}{Robert Nishihara, Iain Murray, and Ryan Adams}

\ShortHeadings{Parallel MCMC with Elliptical Slice Sampling}{Nishihara, Murray, and Adams}
\firstpageno{2087}

\begin{document}

\title{Parallel MCMC with Generalized Elliptical Slice Sampling}

\author{\name Robert Nishihara \email rkn@eecs.berkeley.edu\\
       \addr Department of Electrical Engineering and Computer Science\\
       University of California\\
       Berkeley, CA 94720, USA
       \AND
       \name Iain Murray \email i.murray@ed.ac.uk\\
       \addr School of Informatics\\
       University of Edinburgh\\
       Edinburgh EH8 9AB, UK
       \AND
       \name Ryan P. Adams \email rpa@seas.harvard.edu\\
       \addr School of Engineering and Applied Sciences\\
       Harvard University\\
       Cambridge, MA 02138, USA}

\editor{David Blei}

\maketitle

\begin{abstract}%   <- trailing '%' for backward compatibility of .sty file
Probabilistic models are conceptually powerful tools for finding
structure in data, but their practical effectiveness is often limited
by our ability to perform inference in them. Exact inference is
frequently intractable, so approximate inference is often performed
using Markov chain Monte Carlo (MCMC)\@.  To achieve the best possible
results from MCMC, we want to efficiently simulate many steps of a
rapidly mixing Markov chain which leaves the target distribution
invariant.  Of particular interest in this regard is how to take
advantage of multi-core computing to speed up MCMC-based inference,
both to improve mixing and to distribute the computational load. In
this paper, we present a parallelizable Markov chain Monte Carlo
algorithm for efficiently sampling from continuous probability
distributions that can take advantage of hundreds of cores. This
method shares information between parallel Markov chains to build a
scale-location mixture of Gaussians approximation to the density
function of the target distribution. We combine this approximation
with a recently developed method known as elliptical slice sampling to
create a Markov chain with no step-size parameters that can mix
rapidly without requiring gradient or curvature computations.
\end{abstract}


\section{Introduction}
Probabilistic models are fundamental tools for machine learning,
providing a coherent framework for finding structure in data. In the
Bayesian formulation, learning is performed by computing a
representation of the posterior distribution implied by the
data. Unobserved quantities of interest can then be estimated as
expectations of various functions under this posterior distribution.

These expectations typically correspond to high-dimensional integrals
and sums, which are usually intractable for rich models. There is
therefore significant interest in efficient methods for approximate
inference that can rapidly estimate these expectations. In this paper,
we examine Markov chain Monte Carlo (MCMC) methods for approximate
inference, which estimate these quantities by simulating a Markov
chain with the posterior as its equilibrium distribution. MCMC is
often seen as a principled ``gold standard'' for inference, because
(under mild conditions) its answers will be correct in the limit of
the simulation. However, in practice, MCMC often converges slowly and
requires expert tuning. In this paper, we propose a new method to
address these issues for continuous parameter spaces. We generalize
the method of \emph{elliptical slice sampling} \citep{Murray2010} to
build a new efficient method that: 1)~mixes well in the presence of
strong dependence, 2)~does not require hand tuning, and 3)~can take
advantage of multiple computational cores operating in parallel. We
discuss each of these in more detail below.

Many posterior distributions arising from real-world data have strong
dependencies between variables. These dependencies can arise from
correlations induced by the likelihood function, redundancy in the
parameterization, or directly from the prior. One of the primary
challenges for efficient Markov chain Monte Carlo is making large
moves in directions that reflect the dependence structure. For
example, if we imagine a long, thin region of high density, it is
necessary to explore the length in order to reach equilibrium;
however, random-walk methods such as Metropolis--Hastings (MH) with
spherical proposals can only diffuse as fast as the narrowest
direction allows \citep{Neal1995}. More efficient methods such as
Hamiltonian Monte Carlo \citep{Duane1987, Neal2011, Girolami2011}
avoid random walk behavior by introducing auxiliary ``momentum''
variables. Hamiltonian methods require differentiable density
functions and gradient computations.

In this work, we are able to make efficient long-range moves---even
in the presence of dependence---by building an approximation to the
target density that can be exploited by elliptical slice
sampling. This approximation enables the algorithm to consider the
general shape of the distribution without requiring gradient or
curvature information. In other words, it encodes and allows us to
make use of global information about the distribution as opposed to
the local information used by Hamiltonian Monte Carlo. We construct
the algorithm such that it is valid regardless of the quality of the
approximation, preserving the guarantees of approximate inference by
MCMC\@.

One of the limitations of MCMC in practice is that it is often
difficult for non-experts to apply. This difficulty stems from the
fact that it can be challenging to tune Markov transition operators so
that they mix well. For example, in Metropolis--Hastings, one must
come up with appropriate proposal distributions. In Hamiltonian Monte
Carlo, one must choose the number of steps and the step size in the
simulation of the dynamics. For probabilistic machine learning methods to
be widely applicable, it is necessary to develop black-box methods for
approximate inference that do not require extensive hand tuning. Some
recent attempts have been made in the area of adaptive MCMC
\citep{Roberts2006, Haario2005}, but these are only theoretically
understood for a relatively narrow class of transition operators
(for example, not Hamiltonian Monte Carlo). Here we propose a method based on
slice sampling \citep{Neal2003}, which uses a local search to find an
acceptable point, and avoid potential issues with convergence under
adaptation.

In all aspects of machine learning, a significant challenge is
exploiting a computational landscape that is evolving toward
parallelism over single-core speed. When considering parallel
approaches to MCMC, we can readily identify two ends of a spectrum of
possible solutions. At one extreme, we could run a large number of
independent Markov chains in parallel \citep{Rosenthal2000,
  Bradford1996}. This will have the benefit of providing more samples
and increasing the accuracy of the end result, however it will do
nothing to speed up the convergence or the mixing of each individual
chain. The parallel algorithm will run up against the same limitations
faced by the non-parallel version. At another extreme, we could
develop a single-chain MCMC algorithm which parallelizes the
individual Markov transitions in a problem-specific way. For instance,
if the likelihood is expensive and consists of many factors, the
factors can potentially be computed in parallel. See
\citet{Suchard2010, Tarlow2012a} for examples. Alternatively, some
Markov chain transition operators can make use of multiple parallel
proposals to increase their efficiency, such as multiple-try
Metropolis--Hastings \citep{Liu2000}.

We propose an intermediate algorithm to make effective use of
parallelism. By sharing information between the chains, our method is
able to mix faster than the na\"{i}ve approach of running independent
chains. However, we do not require fine-grained control over parallel
execution, as would be necessary for the single-chain
method. Nevertheless, if such local parallelism is possible, our
sampler can take advantage of it. Our general objective is a black-box
approach that mixes well with multiple cores but does not require the
user to build in parallelism at a low level.

The structure of the paper is as follows. In
Section~\ref{sec:background}, we review slice sampling
\citep{Neal2003} and elliptical slice sampling \citep{Murray2010}. In
Section~\ref{sec:gess}, we show how an elliptical approximation to the
target distribution enables us to generalize elliptical slice sampling
to continuous distributions. In Section~\ref{sec:parallelism}, we
describe a natural way to use parallelism to dynamically construct the
desired approximation. In Section~\ref{sec:related_work}, we discuss
related work. In Section~\ref{sec:experiments}, we evaluate our new
approach against other comparable methods on several typical modeling
problems.

\section{Background}
\label{sec:background}

Throughout this paper, we will use~$\mathcal N({\bf
  x};\boldsymbol\mu,\boldsymbol\Sigma)$ to denote the density function
of a Gaussian with mean~$\boldsymbol\mu$ and
covariance~$\boldsymbol\Sigma$ evaluated at a point~${\bf x} \in
\mathbb R^D$. We will use~$\mathcal
N(\boldsymbol\mu,\boldsymbol\Sigma)$ to refer to the distribution
itself. Analogous notation will be used for other
distributions. Throughout, we shall assume that we wish to draw
samples from a probability distribution over~$\mathbb R^D$ whose
density function is~$\pi$. We sometimes refer to the distribution
itself as~$\pi$.

The objective of Markov chain Monte Carlo is to formulate transition
operators that can be easily simulated, that leave~$\pi$ invariant,
and that are ergodic. Classical examples of MCMC algorithms are
Metropolis--Hastings \citep{Metropolis1953, Hastings1970} and Gibbs
Sampling \citep{Geman1984}.  For general overviews of MCMC, see
\citet{tierney-1994a, andrieu-etal-2003a, HandbookMCMC}. Simulating
such a transition operator will, in the limit, produce samples
from~$\pi$, and these can be used to compute expectations
under~$\pi$. Typically, we only have access to an unnormalized version
of~$\pi$. However, none of the algorithms that we describe require
access to the normalization constant, and so we will abuse notation
somewhat and refer to the unnormalized density as~$\pi$.

\subsection{Slice Sampling}

Slice sampling \citep{Neal2003} is a Markov chain Monte Carlo
algorithm with an adaptive step size. It is an auxiliary-variable
method, which relies on the observation that sampling~$\pi$ is
equivalent to sampling the uniform distribution over the set~$S =
\{({\bf x},y):~0\le~y~\le~\pi({\bf x})\}$ and marginalizing out
the~$y$ coordinate (which in this case is accomplished simply by
disregarding the~$y$ coordinate). Slice sampling accomplishes this by
alternately updating~${\bf x}$ and~$y$ so as to leave invariant the
distributions~$p({\bf x} \given y)$ and~$p(y \given {\bf x})$
respectively. The key insight of slice sampling is that sampling from
these conditionals (which correspond to uniform ``slices'' under the
density function) is potentially much easier than sampling directly
from~$\pi$.

Updating~$y$ according to~$p(y \given {\bf x})$ is trivial. The new
value of~$y$ is drawn uniformly from the interval~$(0, \pi({\bf
  x}))$. There are different ways of updating~${\bf x}$. The objective
is to draw uniformly from among the ``slice''~$\{{\bf x} : \pi({\bf
  x}) \ge y\}$. Typically, this is done by defining a transition
operator that leaves the uniform distribution on the slice
invariant. \citet{Neal2003} describes such a transition operator:
first, choose a direction in which to search, then place an interval
around the current state, expand it as necessary, and shrink it until
an acceptable point is found. Several procedures have been proposed
for the expansion and contraction phases.

Less clear is how to choose an efficient direction in which to
search. There are two approaches that are widely used. First, one
could choose a direction uniformly at random from all possible
directions \citep{MacKay2003}. Second, one could choose a direction
uniformly at random from the~$D$ coordinate directions.  We consider
both of these implementations later, and we refer to them as
\emph{random-direction slice sampling} (RDSS) and
\emph{coordinate-wise slice sampling} (CWSS), respectively. In
principle, any distribution over directions can be used as long as
detailed balance is satisfied, but it is unclear what form this
distribution should take. The choice of direction has a significant
impact on how rapidly mixing occurs. In the remainder of the paper, we
describe how slice sampling can be modified so that candidate points
are chosen to reflect the structure of the target distribution.

\subsection{Elliptical Slice Sampling}

Elliptical slice sampling \citep{Murray2010} is an MCMC algorithm
designed to sample from posteriors over latent variables of the form
\begin{equation} \label{ess}
\pi({\bf x}) \propto L({\bf x}) \, \mathcal N({\bf x};\boldsymbol\mu, \boldsymbol\Sigma),
\end{equation}
where~$L$ is a likelihood function, and~$\mathcal N(\boldsymbol\mu,
\boldsymbol\Sigma)$ is a multivariate Gaussian prior. Such models,
often called \emph{latent Gaussian models}, arise frequently from
Gaussian processes and Gaussian Markov random fields. Elliptical slice
sampling takes advantage of the structure induced by the Gaussian
prior to mix rapidly even when the covariance induces strong
dependence. The method is easier to apply than most MCMC algorithms
because it has no free tuning parameters.

Elliptical slice sampling takes advantage of a convenient invariance
property of the multivariate Gaussian. Namely, if~${\bf x}$
and~${\boldsymbol\nu}$ are independent draws from~$\mathcal
N(\boldsymbol\mu, \boldsymbol\Sigma)$, then the linear combination
\begin{equation}\label{ess2}
{\bf x}' = ({\bf x} - \boldsymbol\mu)\cos \theta  + ({\boldsymbol\nu} - \boldsymbol\mu)\sin \theta + \boldsymbol\mu
\end{equation}
is also (marginally) distributed according to~$\mathcal
N(\boldsymbol\mu, \boldsymbol\Sigma)$ for any~$\theta \in
[0,2\pi]$. Note that~${\bf x'}$ is nevertheless correlated with~${\bf
  x}$ and~${\boldsymbol\nu}$. This correlation has been previously
used to make perturbative Metropolis--Hastings proposals in latent
Gaussian models \citep{Neal1998,Adams2009}, but elliptical slice
sampling uses it as the basis for a rejection-free method.

The elliptical slice sampling transition operator considers the locus
of points defined by varying~$\theta$ in Equation~\eqref{ess2}. This
locus is an ellipse which passes through the current state~${\bf x}$
as well as through the auxiliary variable~$\boldsymbol\nu$. Given a
random ellipse induced by~$\boldsymbol\nu$, we can slice
sample~$\theta \in [0,2\pi]$ to choose the next point based purely on
the likelihood term. The advantage of this procedure is that the
ellipses will necessarily reflect the dependence induced by strong
Gaussian priors and that the user does not have to choose a step
size.

More specifically, elliptical slice sampling updates the current
state~${\bf x}$ as follows. First, the auxiliary
variable~$\boldsymbol\nu \sim \mathcal
N(\boldsymbol\mu,\boldsymbol\Sigma)$ is sampled to define an ellipse
via Equation~\eqref{ess2}, and the value~$u \sim
\textnormal{Uniform}[0,1]$ is sampled to define a likelihood
threshold. Then, a sequence of angles~$\{\theta_k\}$ are chosen
according to a slice-sampling procedure described in
Algorithm~\ref{alg:ess-update}. These angles specify a corresponding
sequence of proposal points~$\{ {\bf x}_k'\}$. We update the current
state~${\bf x}$ by setting it equal to the first proposal point~${\bf
  x}_k'$ satisfying the slice-sampling condition~$L({\bf
  x}_k')>uL({\bf x})$. The proof of the validity of this algorithm is
given in \citet{Murray2010}. Intuitively, the pair~$({\bf
  x},\boldsymbol\nu)$ is updated to a pair~$({\bf
  x}',\boldsymbol\nu')$ with the same joint prior probability, and so
slice sampling only needs to compare likelihood ratios. The new
point~${\bf x}'$ is given by Equation~\eqref{ess2},
while~$\boldsymbol\nu'=(\boldsymbol\nu-\boldsymbol\mu)\cos\theta-({\bf
  x}-\boldsymbol\mu)\sin\theta + \boldsymbol\mu$ is never used and need
not be computed.

\begin{figure}
        \begin{subfigure}[b]{0.5\textwidth}
                \centering
                \includegraphics[scale=1]{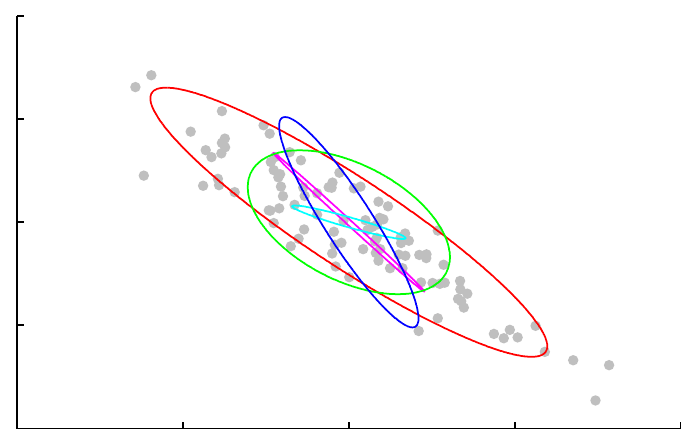}
                \caption{Ellipses from ESS}
        \end{subfigure}
        ~ %add desired spacing between images, e. g. ~, \quad, \qquad etc. 
          %(or a blank line to force the subfigure onto a new line)
        \begin{subfigure}[b]{0.5\textwidth}
                \centering
                \includegraphics[scale=1]{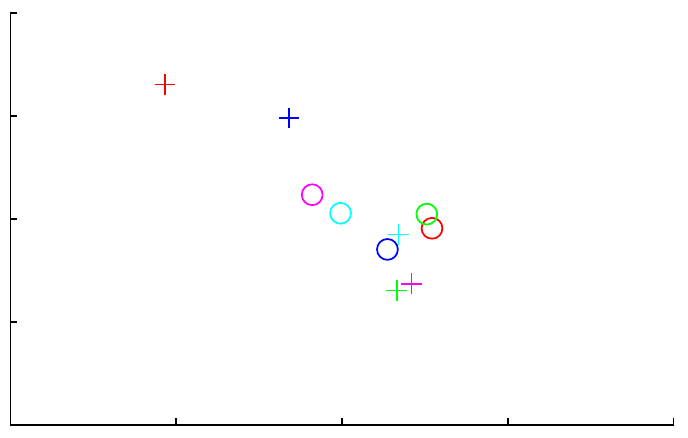}
                \caption{Corresponding values of~${\bf x}$ and~$\boldsymbol\nu$}
        \end{subfigure}%
        \caption{Background points are drawn independently from a probability distribution, and five ellipses are created by elliptical slice sampling. The distribution in question is a two-dimensional multivariate Gaussian. In this example, the same distribution is used as the prior for elliptical slice sampling. {\bf (a)}~Shows the ellipses created by elliptical slice sampling. {\bf (b)}~Shows the values of~${\bf x}$ (depicted as~$\mathlarger{\mathlarger{\mathlarger{\circ}}}$) and~$\boldsymbol\nu$ (depicted as~$\textnormal{+}$) corresponding to each elliptical slice sampling update. The values of~${\bf x}$ and~${\boldsymbol\nu}$ with a given color correspond to the ellipse of the same color in {\bf (a)}.}
        \label{fig:ess_ellipses}
\end{figure}

Figure~\ref{fig:ess_ellipses} depicts random ellipses produced by
elliptical slice sampling superimposed on background points from some
target distribution. This diagram illustrates the idea that the
ellipses produced by elliptical slice sampling reflect the structure
of the distribution. The full algorithm is shown in
Algorithm~\ref{alg:ess-update}.

\begin{algorithm}
\caption{Elliptical Slice Sampling Update}
\label{alg:ess-update}
\begin{algorithmic}[1]
\Require Current state~${\bf x}$, Gaussian parameters~$\boldsymbol\mu$ and~$\boldsymbol\Sigma$, log-likelihood function~$\log L$
\Ensure New state~${\bf x'}$, with stationary distribution proportional to~$\mathcal N({\bf x};\boldsymbol\mu,\boldsymbol\Sigma)L({\bf x})$
\State $\boldsymbol\nu \sim \mathcal N(\boldsymbol\mu, \boldsymbol\Sigma)$ \Comment{Choose ellipse}
\State $u \sim \textnormal{Uniform}[0,1]$
\State $\log y \leftarrow \log L({\bf x}) + \log u$ \Comment{Set log-likelihood threshold}
\State $\theta \sim \textnormal{Uniform}[0,2\pi]$ \Comment{Draw an initial proposal}
\State $[\theta_{\min},\theta_{\max}] \leftarrow [\theta-2\pi, \theta]$ \Comment{Define a bracket}
\State ${\bf x'} \leftarrow ({\bf x} - \boldsymbol\mu)\cos\theta + (\boldsymbol\nu - \boldsymbol\mu) \sin\theta + \boldsymbol\mu$
\If{$\log L({\bf x'}) > \log y$}
  \State {\bf return~${\bf x'}$} \Comment{Accept}
\Else \Comment{Shrink the bracket and try a new point}
  \If{$\theta < 0$}
    \State $\theta_{\min} \leftarrow \theta$
  \Else
    \State $\theta_{\max} \leftarrow \theta$
  \EndIf
  \State $\theta \sim \textnormal{Uniform}[\theta_{\min},\theta_{\max}]$
  \State {\bf goto} 6
\EndIf
\end{algorithmic}
\end{algorithm}

\section{Generalized Elliptical Slice Sampling}
\label{sec:gess}

In this section, we  generalize elliptical slice sampling to handle arbitrary continuous distributions. We refer to this algorithm as \emph{generalized elliptical slice sampling} (GESS)\@. In this section, our target distribution will be a continuous distribution over~$\mathbb R^D$ with density~$\pi$. In practice,~$\pi$ need not be normalized.

Our objective is to reframe our target distribution so that it can be efficiently sampled with elliptical slice sampling. One possible approach is to put~$\pi$ in the form of Equation~\eqref{ess} by choosing some approximation~$\mathcal N(\boldsymbol\mu,\boldsymbol\Sigma)$ to~$\pi$ and writing
\begin{equation*}
  \pi({\bf x}) = R({\bf x}) \, \mathcal N({\bf x};\boldsymbol\mu, \boldsymbol\Sigma),  
\end{equation*}
where
\begin{equation*}
R({\bf x}) = \frac{\pi({\bf x})}{\mathcal N({\bf x};\boldsymbol\mu,\boldsymbol\Sigma)}
\end{equation*}
is the residual error of our approximation to the target density. Note that~$\mathcal N({\bf x};\boldsymbol\mu,\boldsymbol\Sigma)$ is an approximation rather than a prior and that~$R$ is not a likelihood function, but since the equation has the correct form, this representation enables us to use elliptical slice sampling.

\begin{figure}
        \begin{subfigure}[b]{1.0\textwidth}
                \centering
                \includegraphics[scale=1]{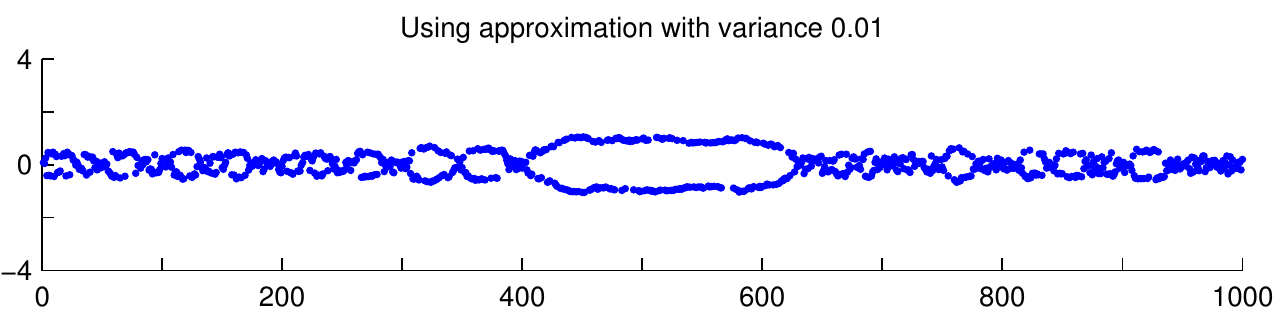}
        \end{subfigure}
        ~

        \begin{subfigure}[b]{1.0\textwidth}
                \centering
                \includegraphics[scale=1]{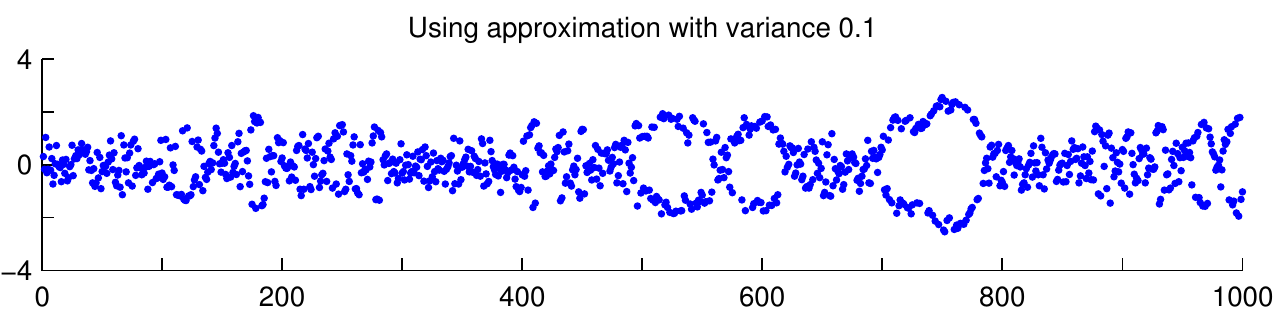}
        \end{subfigure}
        ~

        \begin{subfigure}[b]{1.0\textwidth}
                \centering
                \includegraphics[scale=1]{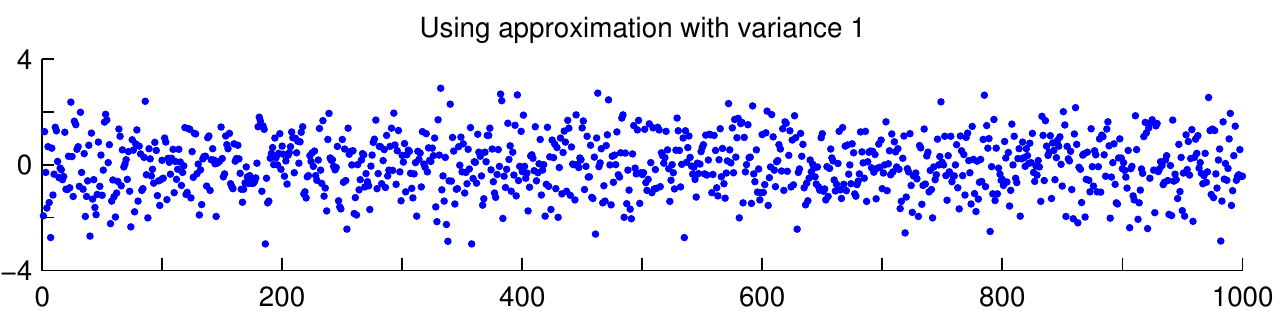}
        \end{subfigure}
        ~

        \begin{subfigure}[b]{1.0\textwidth}
                \centering
                \includegraphics[scale=1]{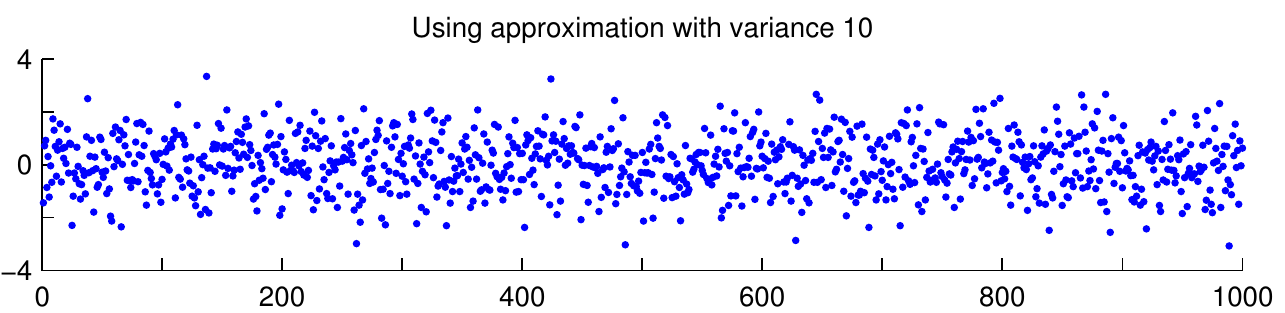}
        \end{subfigure}
        ~ %add desired spacing between images, e. g. ~, \quad, \qquad etc. 
          %(or a blank line to force the subfigure onto a new line)

        \begin{subfigure}[b]{1.0\textwidth}
                \centering
                \includegraphics[scale=1]{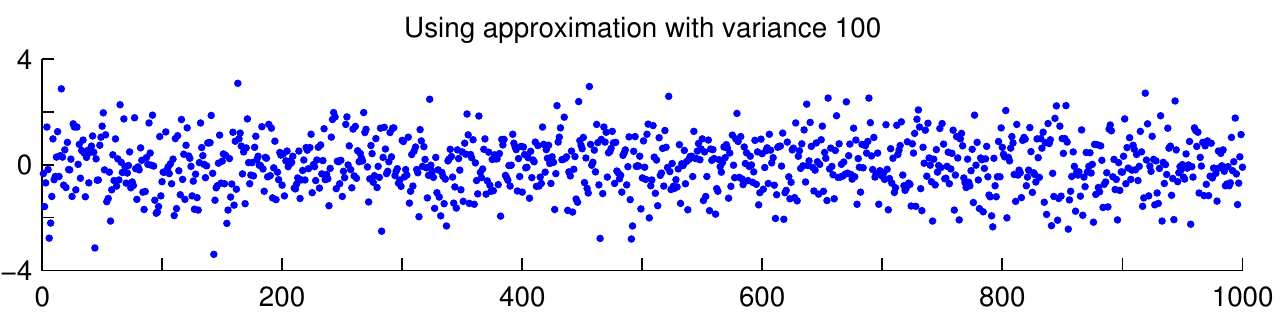}
        \end{subfigure}%
        \caption{Samples are drawn from a Gaussian with zero mean and unit variance using elliptical slice sampling with various Gaussian approximations. These trace plots show how sampling behavior depends on how heavy the tails of the approximation are relative to how heavy the tails of the target distribution are. We plot one of every ten samples.}
        \label{fig:trace_plots}
\end{figure}

Applying elliptical slice sampling in this manner will produce a
correct algorithm, but it may mix slowly in practice. Difficulties
arise when the target distribution has much heavier tails than does
the approximation. In such a situation,~$R({\bf x})$ will increase
rapidly as~${\bf x}$ moves away from the mean of the approximation. To
illustrate this phenomenon, we use this approach with different
approximations to draw samples from a Gaussian in one dimension with
zero mean and unit variance. Trace plots are shown in
Figure~\ref{fig:trace_plots}. The subplot corresponding to
variance~$0.01$ illustrates the problem. Since~$R$ explodes as~$|{\bf
  x}|$ gets large, the Markov chain is unlikely to move back toward
the origin. On the other hand, the size of the ellipse is limited by a
draw from the Gaussian approximation, which has low variance in this
case, so the Markov chain is also unlikely to move away from the
origin. The result is that the Markov chain sometimes gets stuck. In
the subplot corresponding to variance~$0.01$, this occurs between
iterations~$400$ and~$630$.

In order to resolve this pathology and extend elliptical slice
sampling to continuous distributions, we broaden the class of allowed
approximations. To begin with, we express the density of the target
distribution in the more general form
\begin{equation} \label{general-ess}
\pi({\bf x}) \propto R({\bf x}) \int \mathcal N({\bf x} ; \boldsymbol\mu(s), \boldsymbol\Sigma(s)) \, \phi(\mathrm{d}s),
\end{equation}
where the integral represents a scale-location mixture of Gaussians
(which serves as an approximation to~$\pi$), and where~$\phi$ is a
measure over the auxiliary parameter~$s$. As before,~$R$ is the residual
error of the approximation. Here,~$\phi$ can be chosen in a
problem-specific way, and any residual error between~$\pi$ and the
approximation will be compensated for
by~$R$. Equation~\eqref{general-ess} is quite flexible. Below, we will
choose the measure~$\phi$ so as to make the approximation a
multivariate~$t$ distribution, but there are many other
possibilities. For instance, taking~$\phi$ to be a combination of
point masses will make the approximation a discrete mixture of
Gaussians.

Through Equation~\eqref{general-ess}, we can view~$\pi({\bf x})$ as
the marginal density of an augmented joint distribution over~${\bf x}$
and~$s$. Using~$\lambda$ to denote the density of~$\phi$ with respect
to the base measure over~$s$ (this is fully general because we have
control over the choice of base measure), we can write
\begin{equation*} 
p({\bf x}, s) = R({\bf x}) \, \mathcal N({\bf x};\boldsymbol\mu(s),\boldsymbol\Sigma(s)) \, \lambda(s) .
\end{equation*}
Therefore, to sample~$\pi$, it suffices to sample~${\bf x}$ and~$s$
jointly and then to marginalize out the~$s$ coordinate (by simply
dropping the~$s$ coordinate). We update these components alternately
so as to leave invariant the distributions
\begin{equation} \label{x-given-s}
p({\bf x} \given s) \propto R({\bf x}) \, \mathcal N({\bf x};\boldsymbol\mu(s),\boldsymbol\Sigma(s))
\end{equation}
and
\begin{equation} \label{s-given-x}
p(s \given {\bf x}) \propto  \mathcal N({\bf x};\boldsymbol\mu(s),\boldsymbol\Sigma(s)) \, \lambda(s) .
\end{equation}
Equation~\eqref{x-given-s} has the correct form for elliptical slice
sampling and can be updated according to
Algorithm~\ref{alg:ess-update}. Equation~\eqref{s-given-x} can be
updated using any valid Markov transition operator.

We now focus on a particular case in which the update corresponding to
Equation~\eqref{s-given-x} is easy to simulate and in which we can
make the tails as heavy as we desire, so as to control the behavior
of~$R$. A simple and convenient choice is for the scale-location
mixture to yield a multivariate~$t$ distribution with
degrees-of-freedom parameter~$\nu$:
\begin{equation*}
\mathcal T_{\nu}({\bf x};\boldsymbol\mu,\boldsymbol\Sigma) = \int_0^{\infty} \textnormal{IG}(s;\tfrac{\nu}{2},\tfrac{\nu}{2}) \, \mathcal N({\bf x};\boldsymbol\mu,s\boldsymbol\Sigma) \, \mathrm{d}s ,
\end{equation*}
where~$\lambda$ becomes the density function of an inverse-gamma
distribution:
\begin{equation*}
\text{IG}(s;\alpha,\beta) = \frac{\beta^{\alpha}}{\Gamma(\alpha)}s^{-\alpha-1}e^{-\beta/s} .
\end{equation*}
Here~$s$ is a positive real-valued scale parameter. Now, in the
update~$p(s \given {\bf x})$, we observe that the inverse-gamma
distribution acts as a conjugate prior (whose ``prior'' parameters
are~$\alpha=\frac{\nu}{2}$ and~$\beta=\frac{\nu}{2}$), so
\begin{equation*}
p(s\given {\bf x}) = \textnormal{IG}(s;\alpha',\beta')
\end{equation*}
with parameters
\begin{eqnarray}
  \alpha' & = & \frac{D+\nu}{2} \,\,\,\,\,\,\,\, \text{and}  \label{IG-params-alpha} \\ 
  \beta' & = & \frac12(\nu+({\bf x}-\boldsymbol\mu)^{\sf T}\boldsymbol\Sigma^{-1}({\bf x}-\boldsymbol\mu)) .  \label{IG-params-beta}
\end{eqnarray}
We can draw independent samples from this distribution
\citep{Devroye1986}.

\begin{figure}
        \begin{subfigure}[b]{0.5\textwidth}
                \centering
                \includegraphics[scale=1]{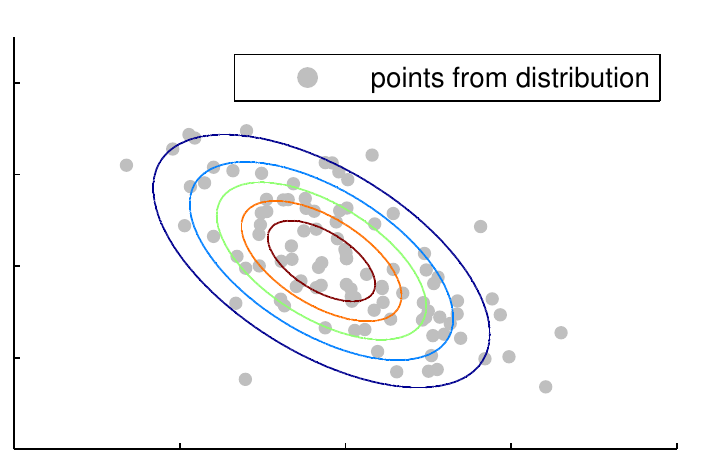}
                \caption{Multivariate~$t$ contours}
        \end{subfigure}
        ~ %add desired spacing between images, e. g. ~, \quad, \qquad etc. 
          %(or a blank line to force the subfigure onto a new line)
        \begin{subfigure}[b]{0.5\textwidth}
                \centering
                \includegraphics[scale=1]{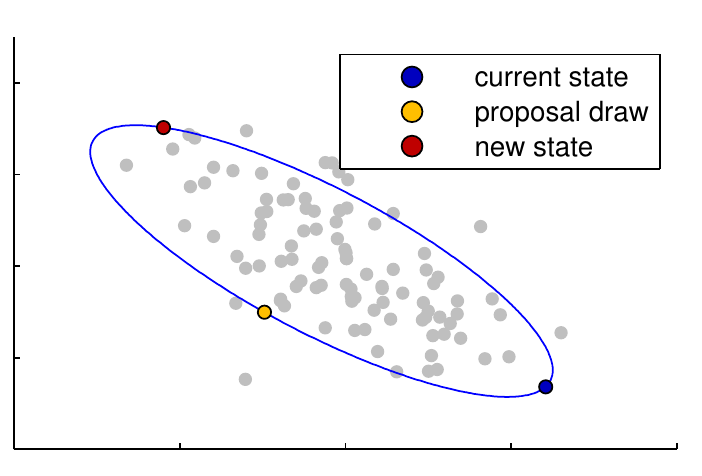}
                \caption{Example GESS update}
        \end{subfigure}%
        \caption{The gray points were drawn independently from a two-dimensional Gaussian to show the mode and shape of the corresponding density function. {\bf (a)}~Shows the contours of a multivariate~$t$ approximation to this distribution. {\bf (b)}~Shows a sample update step using the transition operator~$S({\bf x}\to~{\bf x'};\nu,\boldsymbol\mu,\boldsymbol\Sigma)$. The blue point represents the current state. The yellow point defines an ellipse and is drawn from the Gaussian distribution corresponding to the scale~$s$ drawn from the appropriate inverse-gamma distribution. The red point corresponds to the new state and is sampled from the given ellipse.}
        \label{fig:vis_update}
\end{figure}

Combining these update steps, we define the transition
operator~$S({\bf x} \to {\bf x'}; \nu,
\boldsymbol\mu,\boldsymbol\Sigma)$ to be the one which draws~$s \sim
\textnormal{IG}(s;\alpha',\beta')$, with~$\alpha'$ and~$\beta'$ as
described in Equations~\eqref{IG-params-alpha}
and~\eqref{IG-params-beta}, and then uses elliptical slice sampling to
update~${\bf x}$ so as to leave invariant the distribution defined in
Equation~\eqref{x-given-s}, where~$\boldsymbol\mu(s)=\boldsymbol\mu$
and~$\boldsymbol\Sigma(s)=s\boldsymbol\Sigma$. From the above
discussion, it follows that the stationary distribution of~$S({\bf x}
\to {\bf x'} ;\nu,\boldsymbol\mu,\boldsymbol\Sigma)$
is~$\pi$. Figure~\ref{fig:vis_update} illustrates this transition
operator.

\begin{algorithm}
\caption{Generalized Elliptical Slice Sampling Update}
\label{alg:gess-update}
\begin{algorithmic}[1]
\Require Current state~${\bf x}$, multivariate~$t$ parameters~$\nu, \boldsymbol\mu, \boldsymbol\Sigma$, dimension~$D$, a routine~$\textnormal{ESS}$ that performs an elliptical slice sampling update
\Ensure New state~${\bf x'}$
\State $\alpha' \leftarrow \frac{D+\nu}{2}$
\State $\beta' \leftarrow \frac12(\nu + ({\bf x}-\boldsymbol\mu)^{\sf T}\boldsymbol\Sigma^{-1}({\bf x}-\boldsymbol\mu))$
\State $s \sim \textnormal{IG}(\alpha',\beta')$
\State $\log L \leftarrow \log \pi - \log \mathcal T$ \Comment{$\mathcal T$ is the density of a multivariate~$t$ with parameters~$\nu, \boldsymbol\mu, \boldsymbol\Sigma$}
\State ${\bf x'} \leftarrow \textnormal{ESS}({\bf x}, \boldsymbol\mu, s\boldsymbol\Sigma, \log L)$
\end{algorithmic}
\end{algorithm}

\section{Building the Approximation with Parallelism}
\label{sec:parallelism}

Up to this point, we have not described how to choose the
multivariate~$t$ parameters~$\nu$,~$\boldsymbol\mu$,
and~$\boldsymbol\Sigma$. These choices can be made in many ways. For
instance, we may choose the maximum likelihood parameters given
samples collected during a burn in period, we may build a Laplace
approximation to the mode of the distribution, or we may use
variational approaches. Note that this algorithm is valid regardless
of the particular choice we make here. In this section, we discuss a
convenient way to use parallelism to dynamically choose these
parameters without requiring tuning runs or exploratory analysis of
the distribution. This method creates a large number of parallel
chains, each producing samples from~$\pi$, and it divides them into
two groups. The need for two groups of Markov chains is not
immediately obvious, so we motivate our approach by first discussing
two simpler algorithms that fail in different ways.

\subsection{Na\"ive Approaches}
We begin with a collection of~$K$ parallel chains. Let~${\mathcal
  X~=~\{{\bf x}_1, \ldots, {\bf x}_K\} }$ denote the current states of
the chains. We observe that~$\mathcal X$ may contain a lot of
information about the shape of the target distribution. We would like
to define a transition operator~$Q(\mathcal X \to \mathcal X')$ that
uses this information to intelligently choose the multivariate~$t$
parameters~$\nu$,~$\boldsymbol\mu$, and~$\boldsymbol\Sigma$ and then
uses these parameters to update each~${\bf x}_k$ via generalized
elliptical slice sampling. Additionally, we would like~$Q$ to have two
properties. First, each~${\bf x}_k$ should have the marginal
stationary distribution~$\pi$. Second, we should be able to
parallelize the update of~$\mathcal X$ over~$K$ cores.

Here we describe two simple approaches for parallelizing generalized
elliptical slice sampling, each of which lacks one of the desired
properties. The first approach begins with~$K$ parallel Markov chains,
and it requires a procedure for choosing the multivariate~$t$
parameters given~$\mathcal X$ (for example, maximum likelihood
estimation). In this setup,~$Q$ uses this procedure to determine the
multivariate~$t$ parameters~$\nu_{\mathcal
  X}$,~$\boldsymbol\mu_{\mathcal X}$,~$\boldsymbol\Sigma_{\mathcal X}$
from~$\mathcal X$ and then applies~$S({\bf x} \to {\bf x'};
\nu_{\mathcal X},\boldsymbol\mu_{\mathcal
  X},\boldsymbol\Sigma_{\mathcal X})$ to each~${\bf x}_k$
individually. These updates can be performed in parallel, but the
variables~${\bf x}_k$ no longer have the correct marginal
distributions because of the coupling between the chains introduced by
the approximation (this update violates detailed balance).

A second approach creates a valid MCMC method by including the
multivariate~$t$ parameters in a joint distribution
\begin{equation} \label{eq:naive_parallel}
  p( \mathcal X, \nu, \boldsymbol\mu, \boldsymbol\Sigma) =  p(\nu, \boldsymbol\mu, \boldsymbol\Sigma \given \mathcal X) \left[ \prod_{k=1}^K \pi({\bf x}_k) \right] .
\end{equation}
Note that in Equation~\eqref{eq:naive_parallel}, each~${\bf x}_k$ has
marginal distribution~$\pi$. We can sample this joint distribution by
alternately updating the variables and the multivariate~$t$ parameters
so as to leave invariant the conditional distributions~$p(\mathcal X
\given \nu, \boldsymbol\mu, \boldsymbol\Sigma)$ and~$p(\nu,
\boldsymbol\mu, \boldsymbol\Sigma \given \mathcal X)$. Ideally, we
would like to update the collection~$\mathcal X$ by updating
each~${\bf x}_k$ in parallel. However, we cannot easily parallelize
the update in this formulation because of the factor of~$p(\nu,
\boldsymbol\mu, \boldsymbol\Sigma \given \mathcal X)$, which
nontrivially couples the chains.

\subsection{The Two-Group Approach}

Our proposed method creates a transition operator~$Q$ that satisfies
both of the desired properties. That is, each~${\bf x}_k$ has marginal
distribution~$\pi$, and the update can be efficiently
parallelized. This method circumvents the problems of the previous
approaches by maintaining two groups of Markov chains and using each
group to choose multivariate~$t$ parameters to update the other
group. Let~$\mathcal X = \{{\bf x}_1, \ldots, {\bf x}_{K_1}\}$
and~$\mathcal Y = \{{\bf y}_1, \ldots, {\bf y}_{K_2}\}$ denote the
states of the Markov chains in these two groups (in practice, we set~$K_1=K_2=K$, where~$K$ is the number of available cores). The stationary
distribution of the collection is
\begin{equation*} 
\Pi(\mathcal X, \mathcal Y) = \Pi_1(\mathcal X)\Pi_2(\mathcal Y) = \left[\prod_{k=1}^{K_1} \pi({\bf x}_k)\right] \left[\prod_{k=1}^{K_2} \pi({\bf y}_k)\right] .
\end{equation*}
By simulating a Markov chain which leaves this product distribution invariant, this method generates samples from the target distribution. Our Markov chain is based on a transition operator,~$Q$, defined in two parts. The first part of the transition operator,~$Q_1$, uses~$\mathcal Y$ to determine parameters~$\nu_{\mathcal Y}$,~$\boldsymbol\mu_{\mathcal Y}$, and~$\boldsymbol\Sigma_{\mathcal Y}$. It then uses these parameters to update~$\mathcal X$. The second part of the transition operator,~$Q_2$, uses~$\mathcal X$ to determine parameters~$\nu_{\mathcal X}$,~$\boldsymbol\mu_{\mathcal X}$, and~$\boldsymbol\Sigma_{\mathcal X}$. It then uses these parameters to update~$\mathcal Y$. The transition operator~$Q$ is the composition of~$Q_1$ and~$Q_2$. The idea of maintaining a group of Markov chains and updating the states of some Markov chains based on the states of other Markov chains has been discussed in the literature before. For example, see \citet{Zhang2011, gilks1994}.

In order to make these descriptions more precise, we define~$Q_1$ as follows. First, we specify a procedure for choosing the multivariate~$t$ parameters given the population~$\mathcal Y$. We use an extension of the expectation-maximization algorithm \citep{Liu1995} to choose the maximum-likelihood multivariate~$t$ parameters given the data~$\mathcal Y$. The details of this algorithm are described in Algorithm~\ref{alg:t-params} in the Appendix. More concretely, we choose
\begin{equation*}
 \nu_{\mathcal Y}, \boldsymbol\mu_{\mathcal Y},\boldsymbol\Sigma_{\mathcal Y} = \argmax_{\nu, \boldsymbol\mu,\boldsymbol\Sigma} \prod_{k=1}^{K_2}\mathcal T_{\nu}({\bf y}_{k}\,;\,\boldsymbol\mu,\boldsymbol\Sigma) .
\end{equation*}
After choosing~$\nu_{\mathcal Y}$,~$\boldsymbol\mu_{\mathcal Y}$, and~$\boldsymbol\Sigma_{\mathcal Y}$ in this manner, we update~$\mathcal X$ by applying the transition operator~$S({\bf x} \to~{\bf x'} ;\nu_{\mathcal Y}, \boldsymbol\mu_{\mathcal Y},\boldsymbol\Sigma_{\mathcal Y})$ to each~${\bf x}_k \in \mathcal X$ in parallel. The operator~$Q_2$ is defined analogously.

\begin{algorithm}
\caption{Building the Approximation Using Parallelism}
\label{alg:parallel-update}
\begin{algorithmic}[1]
\Require Two groups of states~$\mathcal X = \{{\bf x}_1, \ldots, {\bf x}_{K_1}\}$ and~$\mathcal Y = \{{\bf y}_1, \ldots, {\bf y}_{K_2}\}$, a subroutine~$\textnormal{FIT-MVT}$ which takes data and returns the maximum-likelihood~$t$ parameters, a subroutine~$\textnormal{GESS}$ which performs a generalized elliptical slice sampling update
\Ensure Updated groups~$\mathcal X'$ and~$\mathcal Y'$
\State $\nu, \boldsymbol\mu, \boldsymbol\Sigma \leftarrow \textnormal{FIT-MVT}(\mathcal Y)$
\ForAll{${\bf x}_k \in \mathcal X$}
  \State ${\bf x}_k' = \textnormal{GESS}({\bf x}_k, \nu, \boldsymbol\mu, \boldsymbol\Sigma)$
\EndFor
\State $\mathcal X' \leftarrow \{{\bf x}_1', \ldots, {\bf x}_{K_1}'\}$
\State $\nu, \boldsymbol\mu, \boldsymbol\Sigma \leftarrow \textnormal{FIT-MVT}(\mathcal X')$
\ForAll{${\bf y}_k \in \mathcal Y$}
  \State ${\bf y}_k' = \textnormal{GESS}({\bf y}_k, \nu, \boldsymbol\mu, \boldsymbol\Sigma)$
\EndFor
\State $\mathcal Y' \leftarrow \{{\bf y}_1', \ldots, {\bf y}_{K_2}'\}$
\end{algorithmic}
\end{algorithm}

In the case where the number of chains in the collection~$\mathcal Y$ is less than or close to the dimension of the distribution, the particular algorithm that we use to choose the parameters \citep{Liu1995} may not converge quickly (or at all). Suppose we are in the setting where~$K<2D$. In this situation, we can perform a regularized estimate of the parameters. We describe this procedure below. The choice~$K<2D$ probably overestimates the regime in which the algorithm for fitting the parameters performs poorly. The particular algorithm that we use appears to work well as long as~$K\ge D$.

Let~$\bar{\bf y}$ be the mean of~$\mathcal Y$, and let~$\{{\bf v}_1, \ldots, {\bf v}_{J}\}$ be the first~${J}$ principal components of the set~$\{{\bf y}_1-\bar{\bf y},\ldots,{\bf y}_K-\bar{\bf y}\}$, where~${J}=\lfloor \tfrac{K}{2}\rfloor$, and let~$V=\textnormal{span}({\bf v}_1,\ldots,{\bf v}_{J})$. Let~${\bf A}$ be the~$D\times {J}$ matrix defined by~${\bf A}{\bf e}_j={\bf v}_j$, where~${\bf e}_j$ is the~$j$th standard basis vector in~$\mathbb R^{J}$. This map identifies~$\mathbb R^{J}$ with~$V$.

Let the set~$\hat{\mathcal Y}$ consist of the projections of the elements of~$\mathcal Y$ onto~$\mathbb R^{J}$ by $\hat{\bf y}_k={\bf A}^{\mathsf T}{\bf y}_k$. Using the algorithm from~\citet{Liu1995}, fit the multivariate~$t$ parameters~$\nu_{\hat{\mathcal Y}}$,~$\boldsymbol\mu_{\hat{\mathcal Y}}$ and,~$\boldsymbol\Sigma_{\hat{\mathcal Y}}$ to~$\hat{\mathcal Y}$. At this point, we have constructed a~${J}$-dimensional multivariate~$t$ distribution, but we would like a~$D$-dimensional one. We construct the desired distribution by rotating back to the original space. Concretely, we can set
\begin{eqnarray*}
  \nu_{\mathcal Y} & = & \nu_{\hat{\mathcal Y}}\\
  \boldsymbol\mu_{\mathcal Y} & = & {\bf A} \, \boldsymbol\mu_{\hat{\mathcal Y}} + \bar{{\bf y}}\\
  \boldsymbol\Sigma_{\mathcal Y} & = & {\bf A} \, \boldsymbol\Sigma_{\hat{\mathcal Y}} \, {\bf A}^{\mathsf T} + \epsilon \, {\bf I}_D,
\end{eqnarray*}
where~${\bf I}_D$ is the~$D\times D$ identity matrix and~$\epsilon$ is the median entry on the diagonal of~$\boldsymbol\Sigma_{\hat{\mathcal Y}}$. We add a scaled identity matrix to the covariance parameter to avoid producing a degenerate distribution. The choice of~$\epsilon$ is based on intuition about typical values of the variance of~$\pi$ in the directions orthogonal to~$V$.

We emphasize that the nature of the procedure for fitting a multivariate~$t$ distribution to some points is not important to our algorithm. One could devise more sophisticated approaches drawing on ideas from the literature on high-dimensional covariance estimation, see \citet{Ravikumar2011} for instance, but we merely choose a simple idea that seems to work in practice. Since our default choice (if there are at least~$2D$ chains, then choose the maximum-likelihood parameters, otherwise project to a lower dimension, choose the maximum-likelihood parameters, and then pad the diagonal of the covariance parameter) works well, the fact that one could design a more sophisticated procedure does not compromise the tuning-free nature of our algorithm.

\subsection{Correctness}

To establish the correctness of our algorithm, we treat the collection
of chains as a single aggregate Markov chain, and we show that this
aggregate Markov chain with transition operator~$Q$ correctly samples
from the product distribution~$\Pi$.

We wish to show that~$Q_1$ and~$Q_2$ preserve the invariant
distributions~$\Pi_1$ and~$\Pi_2$ respectively. As the two cases are
identical, we consider only the first. We have
\begin{eqnarray*}
  \int \Pi_1(\mathcal X) \, Q_1(\mathcal X \to \mathcal X') \, \mathrm{d}\mathcal X & = & \int \Pi_1(\mathcal X) \, Q_1(\mathcal X \to \mathcal X' \given  \nu_{\mathcal Y}, \boldsymbol\mu_{\mathcal Y}, \boldsymbol\Sigma_{\mathcal Y}) \, \mathrm{d}\mathcal X  \\
  & = & \prod_{k=1}^{K_1} \left[ \int \pi({\bf x}_k) \, S({\bf x}_k \to {\bf x}_k'; \nu_{\mathcal Y},\boldsymbol\mu_{\mathcal Y},\boldsymbol\Sigma_{\mathcal Y}) \, \mathrm{d}{\bf x}_k \right] \\
  & = & \Pi_1(\mathcal X') .
\end{eqnarray*}
The last equality uses the fact that~$\pi$ is the stationary
distribution of~$S({\bf x} \to {\bf x'} ;\nu_{\mathcal
  Y},\boldsymbol\mu_{\mathcal Y},\boldsymbol\Sigma_{\mathcal Y})$, so
we see that~$Q$ leaves the desired product distribution invariant.

Within a single chain, elliptical slice sampling has a nonzero
probability of transitioning to any region that has nonzero
probability under the posterior, as described by
\citet{Murray2010}. The transition operator~$Q$ updates the chains in
a given group independently of one another. Therefore~$Q$ has a
nonzero probability of transitioning to any region that has nonzero
probability under the product distribution. It follows that the
transition operator is both irreducible and aperiodic. These
conditions together ensure that this Markov transition operator has a
unique invariant distribution, namely~$\Pi$, and that the
distribution over the state of the Markov chain created from this
transition operator will converge to this invariant distribution
\citep{Roberts2004}. It follows that, in the limit, samples derived
from the repeated application of~$Q$ will be drawn from the desired
distribution.

\subsection{Cost and Complexity}
There is a cost to the construction of the multivariate~$t$
approximation. Although the user has some flexibility in the choice
of~$t$ parameters, we fit them with the iterative algorithm described
by \citet{Liu1995} and in Algorithm~\ref{alg:t-params} of the
Appendix. Let~$D$ be the dimension of the distribution and let~$K$ be
the number of parallel chains. Then the complexity of each iteration
is~$O(D^3K)$, which comes from the fact that we invert a~$D \times D$
matrix for each of the~$K$ chains. Empirically,
Algorithm~\ref{alg:t-params} appears to converge in a small number of
iterations when the number of parallel Markov chains in each group
exceeds the dimension of the distribution. As described in the next
section, this cost can be amortized by reusing the same approximation
for multiple updates. On the challenging distributions that most
interest us, the cost of constructing the approximation (when
amortized in this manner), will be negligible compared to the cost of
evaluating the density function.

An additional concern is the overhead from sharing information between
chains. The chains must communicate in order to build a
multivariate~$t$ approximation, and so the updates must be
synchronized. Since elliptical slice sampling requires a variable
amount of time, updating the different chains will take different
amounts of time, and the faster chains may end up waiting for the
slower ones. We can mitigate this cost by performing multiple updates
between such periods of information sharing. In this manner, we can
perform as much computation as we want between synchronizations
without compromising the validity of the algorithm. As we increase the
number of updates performed between synchronizations, the fraction
of time spent waiting will diminish.

The time measured in our experiments is wall-clock time, which includes the overhead from constructing the approximation and from synchronizing the chains.

\subsection{Reusing the Approximation}
\label{reusing_approx}
Here we explain that reusing the same approximation is valid. To
illustrate this point, let the transition operators~$Q_1$ and~$Q_2$ be
defined as before. In our description of the algorithm, we defined the
transition operator~$Q$ as the composition~${Q = Q_2Q_1}$. However,
both~$Q_1$ and~$Q_2$ preserve the desired product distribution, so we
may use any transition operator of the form~${Q =
  Q_2^{r_2}Q_1^{r_1}}$, where this notation indicates that we first
apply~$Q_1$ for~$r_1$ rounds and then we apply~$Q_2$ for~$r_2$
rounds. As long as~$r_2, r_1 \ge 1$, the composite transition operator
is ergodic. When we apply~$Q_1$ multiple times in a row, the
states~$\mathcal Y$ do not change, so if~$Q_1$ computes~$\nu_{\mathcal
  Y}$,~$\boldsymbol\mu_{\mathcal Y}$, and~$\boldsymbol\Sigma_{\mathcal
  Y}$ deterministically from~$\mathcal Y$, then we need only compute
these values once. Reusing the approximation works even if~$Q_1$
samples~$\nu_{\mathcal Y}$,~$\boldsymbol\mu_{\mathcal Y}$,
and~$\boldsymbol\Sigma_{\mathcal Y}$ from some distribution. In this
case, we can model the randomness by introducing a separate
variable~$r_{\mathcal Y}$ in the Markov chain, and letting~$Q_1$
compute~$\nu_{\mathcal Y}$,~$\boldsymbol\mu_{\mathcal Y}$,
and~$\boldsymbol\Sigma_{\mathcal Y}$ deterministically from~$\mathcal
Y$ and~$r_{\mathcal Y}$.

Our algorithm maintains two collections of Markov chains, one of which
will always be idle. Therefore, each collection can take advantage of
all available cores. Given~$K$ cores, it makes sense to use two
collections of~$K$ Markov chains. In general, it seems to be a good
idea to sample equally from both collections so that the chains in
both collections burn in.

\begin{figure}
        \begin{subfigure}[b]{0.5\textwidth}
                \centering
                \includegraphics[scale=1]{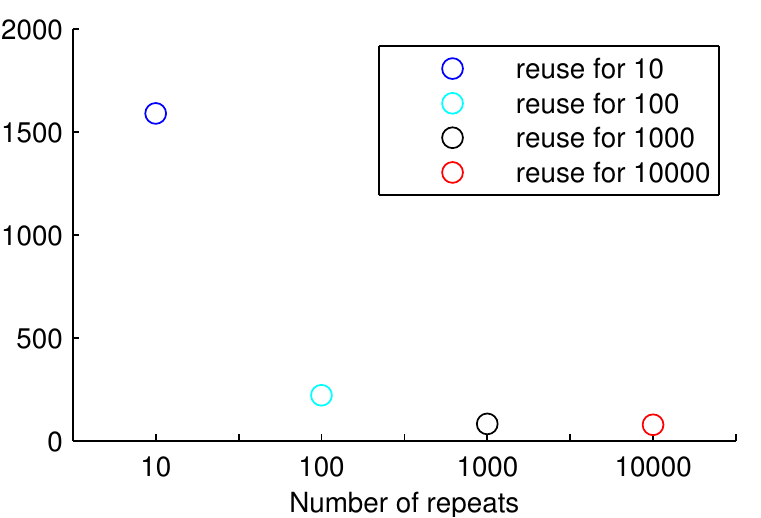}
                \caption{Durations of GESS (seconds)}
        \end{subfigure}
        ~ %add desired spacing between images, e. g. ~, \quad, \qquad etc. 
          %(or a blank line to force the subfigure onto a new line)
        \begin{subfigure}[b]{0.5\textwidth}
                \centering
                \includegraphics[scale=1]{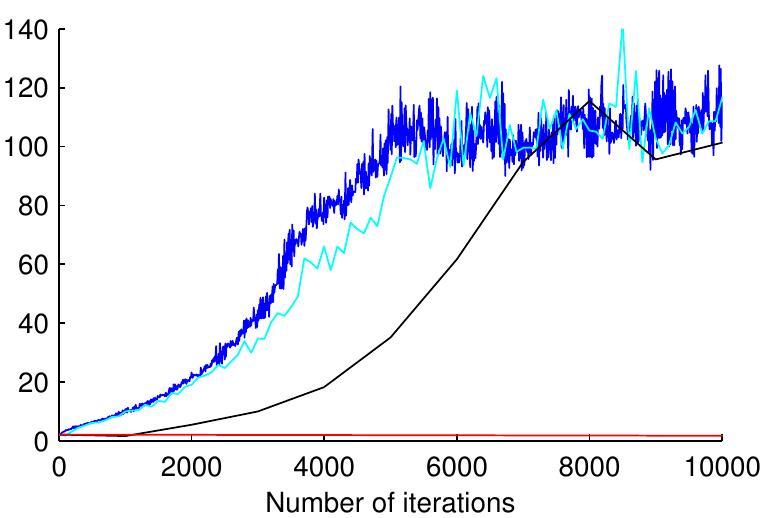}
                \caption{Plots of $\text{tr}(\boldsymbol\Sigma)$}
        \end{subfigure}%
        \caption{We used GESS to sample a multivariate Gaussian distribution in~$100$ dimensions for~$10^4$ iterations. We repeated this procedure four times, reusing the approximation for~$10^1$,~$10^2$,~$10^3$, and~$10^4$ iterations. {\bf (a)}~Shows the durations (in seconds) of the sampling procedures as we varied the number of iterations for which we reused the approximation. {\bf (b)}~Shows how~$\text{tr}(\boldsymbol\Sigma)$ changes over time in the four different settings.}
        \label{fig:reuse_approx}
\end{figure}

To motivate reusing the approximation, we demonstrate the effect of
reusing the approximation for different numbers of iterations on a
Gaussian distribution in~$100$ dimensions (the same one that we use in
Section~\ref{sec:scaling_experiments}). For each value of~$i$ from~$1$
to~$4$, we sample this distribution for~$10^4$ iterations and we reuse
each approximation for~$10^i$ iterations. We show plots of the running
time of GESS and the convergence of the approximation for different
values of~$i$. Figure~\ref{fig:reuse_approx} shows how the amount of
time required by GESS changes as we vary~$i$, and how the covariance
matrix parameter of the fitted multivariate~$t$ approximation changes
over time for the different values of~$i$. We summarize the covariance
matrix parameter by its trace~$\text{tr}(\boldsymbol\Sigma)$. The
figure shows that increasing the number of iterations for which we
reuse the approximation can dramatically reduce the amount of time
required by GESS\@. It also shows that if we rebuild the approximation
frequently, the approximation will settle on a reasonable
approximation in fewer iterations. However, there is little difference
between rebuilding the approximation every~$10$ iterations versus
every~$100$ iterations (in terms of the number of iterations
required), while there is a dramatic difference in the time required.

\section{Related Work}
\label{sec:related_work}

Our work uses updates on a product distribution in the style of
Adaptive Direction Sampling \citep{gilks1994}, which has inspired a
large literature of related methods. The closest research to our work
makes use of slice-sampling based updates of product distributions
along straight-line directions chosen by sampling pairs of points
\citep{MacKay2003,braak2006}. The work on elliptical slice sampling
suggests that in high dimensions larger steps can be taken along
curved trajectories, given an appropriate Gaussian fit. Using closed
ellipses also removes the need to set an initial step size or to build
a bracket.

The recent affine invariant ensemble sampler \citep{goodman2010} also uses
Gaussian fits to a population, in that case to make Metropolis proposals. Our
work differs by using a scale-mixture of Gaussians and elliptical slice sampling
to perform updates on a variety of scales with self-adjusting step-sizes. Rather than updating each member of the population in sequence, our approach splits the population into two groups and allows the members of each group to be updated in parallel.

Population MCMC with parallel tempering \citep{Friel2008} is another
parallel sampling approach that involves sampling from a product
distribution. It uses separate chains to sample a sequence of
distributions interpolating between the target distribution and a
simpler distribution. The different chains regularly swap states to
encourage mixing. In this setting, samples are generated only from a
single chain, and all of the others are auxiliary. However, some tasks
such as computing model evidence can make use of samples from all of
the chains \citep{Friel2008}.

Recent work on Hamiltonian Monte Carlo has attempted to reduce the tuning burden \citep{Hoffman2014}. A user friendly tool that combines this work with  a software stack supporting automatic differentiation is under development \citep{Stan2012}. We feel that this alternative line of work demonstrates the
interest in more practical MCMC algorithms applicable to a variety of
continuous-valued parameter spaces and is very promising. Our complementary
approach introduces simpler algorithms with fewer technical software requirements. In addition, our two-population approach to parallelization could be applied
with whichever methods become dominant in the future.

\section{Experiments}
\label{sec:experiments}

In this section, we compare Algorithm~\ref{alg:parallel-update} with
other parallel MCMC algorithms by measuring how quickly the Markov
chains mix on a number of different distributions. Second, we compare
how the performance of Algorithm~\ref{alg:parallel-update} scales with
the dimension of the target distribution, the number of cores used,
and the number of chains used per core.

These experiments were run on an EC2 cluster with~$5$ nodes, each with
two eight-core Intel Xeon E5-2670 CPUs. We implement all algorithms in
Python, using the IPython environment \citep{Perez2007} for
parallelism.

\subsection{Comparing Mixing} \label{sec:mixing}

We empirically compare the mixing of the parallel MCMC samplers on
seven distributions. We quantify their mixing by comparing the
effective number of samples produced by each method. This quantity can
be approximated as the product of the number of chains with the
effective number of samples from the product distribution. We estimate
the effective number of samples from the product distribution by
computing the effective number of samples from its sequence of log
likelihoods. We compute effective sample size using R-CODA
\citep{coda}, and we compare the results using two metrics: effective
samples per second and effective samples per density function
evaluation (in the case of Hamiltonian Monte Carlo, we count gradient
evaluations as density function evaluations).

In each experiment, we run each algorithm with~$100$ parallel
chains. Unless otherwise noted, we burn in for~$10^4$ iterations and
sample for~$10^5$ iterations. We run five trials for each experiment
to estimate variability.

Figure~\ref{fig:results} shows the average effective number of
samples, with error bars, according to the two different metrics. Bars
are omitted where the sequence of aggregate log likelihoods did not
converge according to the Geweke convergence diagnostic
\citep{Geweke1992}. We diagnose this using the tool from R-CODA
\citep{coda}.

\subsubsection{Samplers Considered}
We compare generalized elliptical slice sampling (GESS) with parallel
versions of several different sampling algorithms.

First, we consider random-direction slice sampling (RDSS)
\citep{MacKay2003} and coordinate-wise slice sampling (CWSS)
\citep{Neal2003}. These are variants of slice sampling which differ in
their choice of direction (a random direction versus a random
axis-aligned direction) in which to sample. RDSS is rotation invariant
like GESS, but CWSS is not.

In addition, we compare to a simple Metropolis--Hastings (MH)
\citep{Metropolis1953} algorithm whose proposal distribution is a
spherical Gaussian centered on the current state. A tuning period is
used to adjust the MH step size so that the acceptance ratio is as
close as possible to the value~$0.234$, which is optimal in some
settings \citep{Roberts1998}. This tuning is done independently for
each chain. We also compare to an adaptive MCMC (AMH) algorithm
following the approach in \citet{Roberts2006} in which the covariance
of a Metropolis--Hastings proposal is adapted to the history of the
``Markov'' chain.

We also compare to the No-U-Turn sampler \citep{Hoffman2014}, which is
an implementation of Hamiltonian Monte Carlo (HMC) combined with
procedures to automatically tune the step size parameter and the
number of steps parameter. Due to the large number of function
evaluations per sample required by HMC, we run HMC for a factor
of~$10$ or~$100$ fewer iterations in order to make the algorithms
roughly comparable in terms of wall-clock time. Though we include the
comparisons, we do not view HMC as a perfectly comparable algorithm
due to its requirement that the density function of the target
distribution be differentiable. Though the target distribution is
often differentiable in principle, there are many practical situations
in which the gradient is difficult to access, either by manual
computation or by automatic differentiation, possibly because
evaluating the density function requires running a complicated
black-box subroutine. For instance, in computer vision problems,
evaluating the likelihood function may require rendering an image or
running graph cuts. See \citet{Tarlow2012b} or \citet{Lang2012} for
examples.

We compare to parallel tempering (PT) \citep{Friel2008}, using each
Markov chain to sample the distribution at a different temperature (if
the target distribution has density~$\pi({\bf x})$, then the
distribution ``at temperature~$t$'' has density proportional
to~$\pi({\bf x})^{1/t}$) and swapping states between the Markov chains
at regular intervals. Samples from the target distribution are
produced by only one of the chains. Using PT requires the practitioner
to pick a temperature schedule, and doing so often requires a
significant amount of experimentation \citep{Neal2001}. We follow the
practice of \citet{Friel2008} and use a geometric temperature
schedule. As with HMC, we do not view PT as entirely comparable in the
absence of an automatic and principled way to choose the temperatures
of the different Markov chains. One of the main goals of GESS is to
provide a black-box MCMC algorithm that imposes as few restrictions on
the target distribution as possible and that requires no expertise or
experimentation on the part of the user.

\subsubsection{Distributions}
In this section, we describe the different distributions that we use to compare the mixing of our samplers.

%\paragraph{Funnel:}
{\em Funnel:} A ten-dimensional funnel-shaped distribution described in
\citet{Neal2003}. The first coordinate is distributed normally with
mean zero and variance nine. Conditioned on the first coordinate~$v$,
the remaining coordinates are independent identically-distributed
normal random variables with mean zero and variance~$e^v$. In this
experiment, we initialize each Markov chain from a spherical
multivariate Gaussian centered on the origin.

%\paragraph{Gaussian Mixture:}
{\em Gaussian Mixture:} An eight-component mixture of Gaussians in eight dimensions. Each
component is a spherical Gaussian with unit variance. The components
are distributed uniformly at random within a hypercube of edge length
four. In this experiment, we initialize each Markov chain from a
spherical multivariate Gaussian centered on the origin.

%\paragraph{Breast Cancer:}
{\em Breast Cancer:} The posterior density of a linear logistic regression model for a
binary classification problem with thirty explanatory variables
(thirty-one dimensions) using the Breast Cancer Wisconsin data set
\citep{Street1993}. The data set consists of~$569$ data points. We
scale the data so that each coordinate has unit variance, and we place
zero-mean Gaussian priors with variance~$100$ on each of the
regression coefficients. In this experiment, we initialize each Markov
chain from a spherical multivariate Gaussian centered on the origin.

%\paragraph{German Credit:}
{\em German Credit:} The posterior density of a linear logistic regression model for a
binary classification problem with twenty-four explanatory variables
(twenty-five dimensions) from the UCI repository
\citep{Frank2010}. The data set consists of~$1000$ data points. We
scale the data so that each coordinate has unit variance, and we place
zero-mean Gaussian priors with variance~$100$ on each of the
regression coefficients. In this experiment, we initialize each Markov
chain from a spherical multivariate Gaussian centered on the origin.

%\paragraph{Stochastic Volatility:}
{\em Stochastic Volatility:} The posterior density of a simple stochastic volatility model fit to
synthetic data in fifty-one dimensions. This distribution is a smaller
version of a distribution described in \citet{Hoffman2014}. In this
experiment, we burn-in for~$10^5$ iterations and sample for~$10^5$
iterations. We initialize each Markov chain from a spherical
multivariate Gaussian centered on the origin and we take the absolute
value of the first parameter, which is constrained to be positive.

%\paragraph{Ionosphere:}
{\em Ionosphere:} The posterior density on covariance hyperparameters for Gaussian
process regression applied to the Ionosphere data set
\citep{Sigillito1989}. We use a squared exponential kernel with
thirty-four length-scale hyperparameters and~$100$ data points. We
place exponential priors with rate~$0.1$ on the length-scale
hyperparameters. In this experiment, we burn-in for~$10^4$ iterations
and sample for~$10^4$ iterations. We initialize each Markov chain from
a spherical multivariate Gaussian centered on the vector~$(1, \ldots,
1)^{\mathsf T}$.

%\paragraph{SNP Covariates:}
{\em SNP Covariates:} The posterior density of the parameters of a generative model for gene
expression levels simulated in thirty-eight dimensions using actual
genomic sequences from~$480$ individuals for covariate data
\citep{Engelhardt2014}. In this experiment, we burn-in for~$2000$
iterations and sample for~$10^4$ iterations. We initialize each Markov
chain from a spherical multivariate Gaussian centered on the origin
and we take the absolute value of the first three parameters, which
are constrained to be positive.

\subsubsection{Mixing Results}

\begin{figure}
        \begin{subfigure}[b]{0.5\textwidth}
                \centering
                \includegraphics[scale=1]{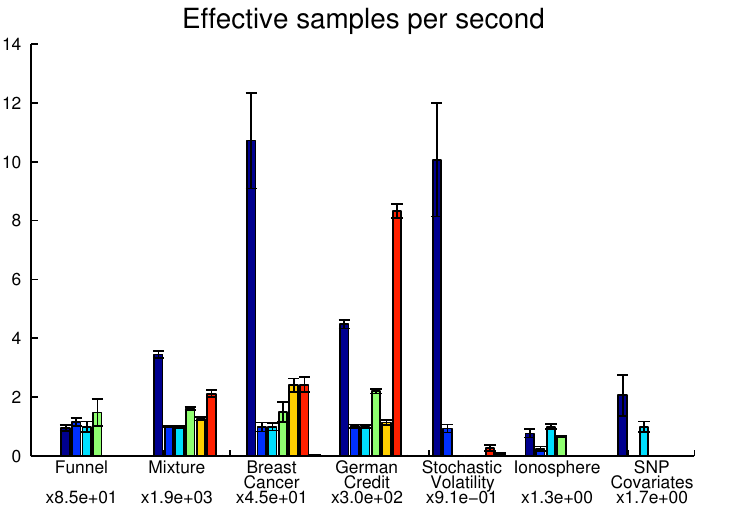}
        \end{subfigure}%
        ~ %add desired spacing between images, e. g. ~, \quad, \qquad etc. 
          %(or a blank line to force the subfigure onto a new line)
        \begin{subfigure}[b]{0.5\textwidth}
                \centering
                \includegraphics[scale=1]{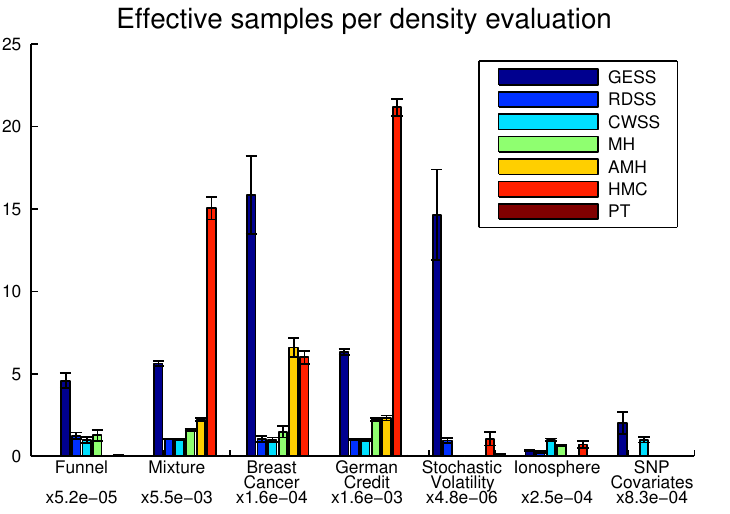}
        \end{subfigure}
        \caption{The results of experimental comparisons of seven parallel MCMC methods on seven distributions. Each figure shows seven groups of bars, (one for each distribution) and the vertical axis shows the effective number of samples per unit cost. Error bars are included. Bars are omitted where the method failed to converge according to the Geweke diagnostic \citep{Geweke1992}. The costs are \emph{per second} (left) and \emph{per density function evaluation} (right). Mean and standard error for five runs are shown. Each group of bars has been rescaled for readability: the number beneath each group gives the effective samples corresponding to CWSS, which always has height one. }
        \label{fig:results}
\end{figure}

The results of the mixing experiments are shown in
Figure~\ref{fig:results}. For the most part, GESS sampled more
effectively than the other algorithms according to both metrics. The
poor performance of PT can be attributed to the fact that PT only
produces samples from one of its chains, unlike the other algorithms,
which produce samples from~$100$ chains. HMC also performed well,
although it failed to converge on the SNP Covariates distribution. The
density function of this particular distribution is only piecewise
continuous, with the discontinuities arising from thresholding in the
model. In this case, the gradient and curvature largely reflect the
prior, whereas the likelihood mostly manifests itself in the
discontinuities of the distribution.

One reason for the rapid mixing of GESS is that GESS performs well
even on highly-skewed distributions. RDSS, CWSS, MH, and PT propose
steps in uninformed directions, the vast majority of which lead away
from the region of high density. As a result, these algorithms take
very small steps, causing successive states to be highly
correlated. In the case of GESS, the multivariate~$t$ approximation
builds information about the global shape of the distribution
(including skew) into the transition operator. As a consequence, the
Markov chain can take long steps along the length of the distribution,
allowing the Markov chain to mix much more rapidly. Skewed
distributions can arise as a result of the user not knowing the
relative length scales of the parameters or as a result of redundancy
in the parameterization. Therefore, the ability to perform well on
such distributions is frequently relevant.

These results show that a multivariate~$t$ approximation to the target
distribution provides enough information to greatly speed up the
mixing of the sampler and that this information can be used to improve
the convergence of the sampler. These improvements occur on top of the
performance gain from using parallelism.

\subsection{Scaling the Number of Cores} \label{sec:scaling_experiments}

\begin{figure}
  \begin{subfigure}[b]{1.0\textwidth}
    \centering
  \begin{tabularx}{\textwidth}{ X|XXXXX }
    $D=50$ & $K=C$ & $K=2C$ & $K=3C$ & $K=4C$ & $K=5C$ \\
    \hline
$C=20$ & ${\color{blue}\scriptstyle{10^{-0.5} \pm 10^{-0.4}}}$ & ${\color{blue}\scriptstyle{10^{-1.2} \pm 10^{-1.5}}}$ & ${\color{blue}\scriptstyle{10^{-1.5} \pm 10^{-1.8}}}$ & ${\color{blue}\scriptstyle{10^{-1.7} \pm 10^{-1.8}}}$ & ${\color{blue}\scriptstyle{10^{-1.6} \pm 10^{-1.6}}}$ \\
$C=40$ & ${\color{blue}\scriptstyle{10^{-0.8} \pm 10^{-0.9}}}$ & ${\color{blue}\scriptstyle{10^{-2.6} \pm 10^{-2.6}}}$ & ${\color{blue}\scriptstyle{10^{-1.9} \pm 10^{-1.9}}}$ & ${\color{blue}\scriptstyle{10^{-1.8} \pm 10^{-1.8}}}$ & ${\color{blue}\scriptstyle{10^{-2.4} \pm 10^{-2.6}}}$ \\
$C=60$ & ${\color{blue}\scriptstyle{10^{-1.6} \pm 10^{-1.5}}}$ & ${\color{blue}\scriptstyle{10^{-1.6} \pm 10^{-1.7}}}$ & ${\color{blue}\scriptstyle{10^{-2.1} \pm 10^{-2.2}}}$ & ${\color{blue}\scriptstyle{10^{-2.1} \pm 10^{-2.2}}}$ & ${\color{blue}\scriptstyle{10^{-2.2} \pm 10^{-2.4}}}$ \\
$C=80$ & ${\color{blue}\scriptstyle{10^{-1.3} \pm 10^{-1.1}}}$ & ${\color{blue}\scriptstyle{10^{-2.4} \pm 10^{-2.8}}}$ & ${\color{blue}\scriptstyle{10^{-2.4} \pm 10^{-2.4}}}$ & ${\color{blue}\scriptstyle{10^{-2.1} \pm 10^{-2.4}}}$ & ${\color{blue}\scriptstyle{10^{-2.3} \pm 10^{-2.5}}}$ \\
$C=100$ & ${\color{blue}\scriptstyle{10^{-1.6} \pm 10^{-1.7}}}$ & ${\color{blue}\scriptstyle{10^{-1.7} \pm 10^{-1.7}}}$ & ${\color{blue}\scriptstyle{10^{-2.0} \pm 10^{-2.0}}}$ & ${\color{blue}\scriptstyle{10^{-2.2} \pm 10^{-2.4}}}$ & ${\color{blue}\scriptstyle{10^{-2.5} \pm 10^{-2.3}}}$
  \end{tabularx}
  %\caption{Error in the estimate of $\boldsymbol\sigma$ for $D=50$.}
  \end{subfigure}
  \vspace{3.5em}

  \begin{subfigure}[b]{1.0\textwidth}
    \centering
  \begin{tabularx}{\textwidth}{ X|XXXXX }
    $D=100$ & $K=C$ & $K=2C$ & $K=3C$ & $K=4C$ & $K=5C$ \\
    \hline
$C=20$ & $\scriptstyle{10^{+0.3} \pm 10^{+0.2}}$ & ${\color{blue}\scriptstyle{10^{-1.3} \pm 10^{-2.2}}}$ & ${\color{blue}\scriptstyle{10^{-1.7} \pm 10^{-2.1}}}$ & ${\color{blue}\scriptstyle{10^{-1.9} \pm 10^{-2.2}}}$ & ${\color{blue}\scriptstyle{10^{-2.4} \pm 10^{-3.5}}}$ \\
$C=40$ & ${\color{blue}\scriptstyle{10^{-1.1} \pm 10^{-1.1}}}$ & ${\color{blue}\scriptstyle{10^{-1.9} \pm 10^{-2.1}}}$ & ${\color{blue}\scriptstyle{10^{-2.5} \pm 10^{-3.2}}}$ & ${\color{blue}\scriptstyle{10^{-2.5} \pm 10^{-2.6}}}$ & ${\color{blue}\scriptstyle{10^{-2.7} \pm 10^{-3.0}}}$ \\
$C=60$ & ${\color{blue}\scriptstyle{10^{-1.7} \pm 10^{-2.0}}}$ & ${\color{blue}\scriptstyle{10^{-2.5} \pm 10^{-2.8}}}$ & ${\color{blue}\scriptstyle{10^{-2.8} \pm 10^{-3.0}}}$ & ${\color{blue}\scriptstyle{10^{-2.9} \pm 10^{-3.4}}}$ & ${\color{blue}\scriptstyle{10^{-2.9} \pm 10^{-3.1}}}$ \\
$C=80$ & ${\color{blue}\scriptstyle{10^{-2.1} \pm 10^{-2.7}}}$ & ${\color{blue}\scriptstyle{10^{-2.7} \pm 10^{-2.8}}}$ & ${\color{blue}\scriptstyle{10^{-2.7} \pm 10^{-3.0}}}$ & ${\color{blue}\scriptstyle{10^{-2.9} \pm 10^{-3.2}}}$ & ${\color{blue}\scriptstyle{10^{-3.1} \pm 10^{-4.0}}}$ \\
$C=100$ & ${\color{blue}\scriptstyle{10^{-2.4} \pm 10^{-2.6}}}$ & ${\color{blue}\scriptstyle{10^{-2.8} \pm 10^{-3.3}}}$ & ${\color{blue}\scriptstyle{10^{-3.0} \pm 10^{-3.5}}}$ & ${\color{blue}\scriptstyle{10^{-3.0} \pm 10^{-3.6}}}$ & ${\color{blue}\scriptstyle{10^{-2.9} \pm 10^{-3.0}}}$
  \end{tabularx}
  %\caption{Error in the estimate of $\boldsymbol\sigma$ for $D=100$.}
  \end{subfigure}
  \vspace{3.5em}

  \begin{subfigure}[b]{1.0\textwidth}
    \centering
  \begin{tabularx}{\textwidth}{ X|XXXXX }
    $D=150$ & $K=C$ & $K=2C$ & $K=3C$ & $K=4C$ & $K=5C$ \\
    \hline
$C=20$ & $\scriptstyle{10^{+2.3} \pm 10^{+1.4}}$ & $\scriptstyle{10^{+2.3} \pm 10^{+1.7}}$ & $\scriptstyle{10^{+1.4} \pm 10^{+1.0}}$ & $\scriptstyle{10^{+0.5} \pm 10^{+0.2}}$ & ${\color{blue}\scriptstyle{10^{-0.7} \pm 10^{-1.0}}}$ \\
$C=40$ & $\scriptstyle{10^{+2.1} \pm 10^{+1.6}}$ & ${\color{blue}\scriptstyle{10^{-0.1} \pm 10^{-0.0}}}$ & ${\color{blue}\scriptstyle{10^{-1.1} \pm 10^{-1.2}}}$ & ${\color{blue}\scriptstyle{10^{-1.4} \pm 10^{-1.4}}}$ & ${\color{blue}\scriptstyle{10^{-1.8} \pm 10^{-1.7}}}$ \\
$C=60$ & $\scriptstyle{10^{+1.3} \pm 10^{+0.7}}$ & ${\color{blue}\scriptstyle{10^{-1.2} \pm 10^{-1.2}}}$ & ${\color{blue}\scriptstyle{10^{-1.6} \pm 10^{-1.5}}}$ & ${\color{blue}\scriptstyle{10^{-1.9} \pm 10^{-2.0}}}$ & ${\color{blue}\scriptstyle{10^{-1.7} \pm 10^{-1.6}}}$ \\
$C=80$ & ${\color{blue}\scriptstyle{10^{-0.0} \pm 10^{-0.0}}}$ & ${\color{blue}\scriptstyle{10^{-1.7} \pm 10^{-1.8}}}$ & ${\color{blue}\scriptstyle{10^{-2.2} \pm 10^{-2.3}}}$ & ${\color{blue}\scriptstyle{10^{-1.9} \pm 10^{-2.0}}}$ & ${\color{blue}\scriptstyle{10^{-2.1} \pm 10^{-2.6}}}$ \\
$C=100$ & ${\color{blue}\scriptstyle{10^{-0.7} \pm 10^{-1.0}}}$ & ${\color{blue}\scriptstyle{10^{-1.8} \pm 10^{-2.1}}}$ & ${\color{blue}\scriptstyle{10^{-1.9} \pm 10^{-2.1}}}$ & ${\color{blue}\scriptstyle{10^{-2.0} \pm 10^{-2.1}}}$ & ${\color{blue}\scriptstyle{10^{-2.3} \pm 10^{-2.3}}}$
  \end{tabularx}
  %\caption{Error in the estimate of $\boldsymbol\sigma$ for $D=150$.}
  \end{subfigure}
  \vspace{3.5em}

  \begin{subfigure}[b]{1.0\textwidth}
    \centering
  \begin{tabularx}{\textwidth}{ X|XXXXX }
    $D=200$ & $K=C$ & $K=2C$ & $K=3C$ & $K=4C$ & $K=5C$ \\
    \hline
$C=20$ & $\scriptstyle{10^{+2.8} \pm 10^{+2.5}}$ & $\scriptstyle{10^{+3.0} \pm 10^{+2.4}}$ & $\scriptstyle{10^{+3.1} \pm 10^{+2.1}}$ & $\scriptstyle{10^{+3.1} \pm 10^{+1.9}}$ & $\scriptstyle{10^{+3.0} \pm 10^{+1.5}}$ \\
$C=40$ & $\scriptstyle{10^{+3.1} \pm 10^{+1.6}}$ & $\scriptstyle{10^{+3.1} \pm 10^{+1.7}}$ & $\scriptstyle{10^{+2.7} \pm 10^{+1.6}}$ & $\scriptstyle{10^{+1.1} \pm 10^{+0.6}}$ & ${\color{blue}\scriptstyle{10^{-1.4} \pm 10^{-1.6}}}$ \\
$C=60$ & $\scriptstyle{10^{+3.1} \pm 10^{+1.6}}$ & $\scriptstyle{10^{+2.6} \pm 10^{+1.8}}$ & ${\color{blue}\scriptstyle{10^{-0.6} \pm 10^{-0.8}}}$ & ${\color{blue}\scriptstyle{10^{-1.7} \pm 10^{-2.0}}}$ & ${\color{blue}\scriptstyle{10^{-2.0} \pm 10^{-2.8}}}$ \\
$C=80$ & $\scriptstyle{10^{+3.1} \pm 10^{+1.7}}$ & $\scriptstyle{10^{+0.7} \pm 10^{+0.1}}$ & ${\color{blue}\scriptstyle{10^{-1.7} \pm 10^{-2.3}}}$ & ${\color{blue}\scriptstyle{10^{-1.9} \pm 10^{-1.9}}}$ & ${\color{blue}\scriptstyle{10^{-2.1} \pm 10^{-2.5}}}$ \\
$C=100$ & $\scriptstyle{10^{+3.0} \pm 10^{+2.1}}$ & ${\color{blue}\scriptstyle{10^{-1.4} \pm 10^{-1.6}}}$ & ${\color{blue}\scriptstyle{10^{-2.3} \pm 10^{-2.8}}}$ & ${\color{blue}\scriptstyle{10^{-2.0} \pm 10^{-2.6}}}$ & ${\color{blue}\scriptstyle{10^{-2.3} \pm 10^{-2.9}}}$
  \end{tabularx}
  %\caption{Error in the estimate of $\boldsymbol\sigma$ for $D=200$.}
  \end{subfigure}
  %add desired spacing between images, e. g. ~, \quad, \qquad etc. 
  %(or a blank line to force the subfigure onto a new line)
  \vspace{3.5em}

  \begin{subfigure}[b]{1.0\textwidth}
    \centering
  \begin{tabularx}{\textwidth}{ X|XXXXX }
    $D=250$ & $K=C$ & $K=2C$ & $K=3C$ & $K=4C$ & $K=5C$ \\
    \hline
$C=20$ & $\scriptstyle{10^{+3.5} \pm 10^{+2.0}}$ & $\scriptstyle{10^{+3.5} \pm 10^{+1.5}}$ & $\scriptstyle{10^{+3.5} \pm 10^{+1.7}}$ & $\scriptstyle{10^{+3.5} \pm 10^{+1.4}}$ & $\scriptstyle{10^{+3.5} \pm 10^{+1.6}}$ \\
$C=40$ & $\scriptstyle{10^{+3.5} \pm 10^{+2.3}}$ & $\scriptstyle{10^{+3.5} \pm 10^{+1.3}}$ & $\scriptstyle{10^{+3.5} \pm 10^{+1.6}}$ & $\scriptstyle{10^{+3.5} \pm 10^{+2.1}}$ & $\scriptstyle{10^{+3.6} \pm 10^{+1.8}}$ \\
$C=60$ & $\scriptstyle{10^{+3.5} \pm 10^{+2.0}}$ & $\scriptstyle{10^{+3.5} \pm 10^{+1.6}}$ & $\scriptstyle{10^{+3.6} \pm 10^{+2.1}}$ & $\scriptstyle{10^{+3.6} \pm 10^{+2.4}}$ & $\scriptstyle{10^{+2.3} \pm 10^{+1.9}}$ \\
$C=80$ & $\scriptstyle{10^{+3.5} \pm 10^{+1.6}}$ & $\scriptstyle{10^{+3.5} \pm 10^{+1.9}}$ & $\scriptstyle{10^{+3.5} \pm 10^{+2.2}}$ & $\scriptstyle{10^{+1.1} \pm 10^{+0.8}}$ & ${\color{blue}\scriptstyle{10^{-0.8} \pm 10^{-0.9}}}$ \\
$C=100$ & $\scriptstyle{10^{+3.5} \pm 10^{+1.8}}$ & $\scriptstyle{10^{+3.6} \pm 10^{+2.0}}$ & $\scriptstyle{10^{+2.2} \pm 10^{+1.7}}$ & $\scriptstyle{10^{+0.3} \pm 10^{+0.2}}$ & ${\color{blue}\scriptstyle{10^{-0.1} \pm 10^{-0.2}}}$
  \end{tabularx}
  %\caption{Error in the estimate of $\boldsymbol\sigma$ for $D=250$.}
  \end{subfigure}%
  \caption{For each choice of~$D$,~$C$, and~$K$, we run GESS, estimate~$\boldsymbol\sigma$, and report the squared error averaged over~$5$ trials along with error bars. Smaller numbers are better. Average errors less than~$1$ are shown in blue.}
  \label{fig:scaling_table}
\end{figure}

We wish to explore the performance of GESS as a function of the
dimension~$D$ of the target distribution, the number~$C$ of cores
available, and the number~$K$ of parallel chains. In this experiment,
we consider all~$125$ triples~$(D,C,K)$ such that
\begin{eqnarray*}
  D & \in & \{50,100,150,200,250\} \\
  C & \in & \{20,40,60,80,100\} \\
  K & \in & \{C,2C,3C,4C,5C\} .
\end{eqnarray*}
It makes sense to let~$K$ be an integer multiple of~$C$ so that each
core will be tasked with updating the same number of chains (the
experiments in Section~\ref{sec:mixing} set~$K$ equal to~$C$).

For each triple~$(D,C,K)$, we sample a~$D$-dimensional multivariate
Gaussian distribution centered on the origin whose precision matrix
was generated from a Wishart distribution with identity scale matrix
and~$D$ degrees of freedom. The distributions used in this experiment
were modeled off of one of the distributions considered
in~\citet{Hoffman2014}. We initialize GESS from a broad spherical
Gaussian distribution centered on the origin, and we run GESS
for~$500$ seconds. The first half of the resulting samples are
discarded, and the second half of the resulting samples are used to
estimate the vector~$\boldsymbol\sigma=(\sigma_1,\ldots,\sigma_D)$,
where~$\sigma_d$ is the marginal standard deviation of the~$d$th
coordinate. For each triple~$(D,C,K)$, we run five
trials. Figure~\ref{fig:scaling_table} shows the resulting average
squared error in the empirical estimate of~$\boldsymbol\sigma$ after
$500$ seconds. Error bars are included as well.

When aggregating samples from~$K$ independent Markov chains, we would
expect the squared error of our estimator to decrease at the
rate~$1/K$. However, in the setting of GESS, additional chains not
only provide additional samples, but may enable the construction of a
more accurate approximation to the target distribution thereby
enabling the other chains to sample more effectively. In some
situations, the presence of additional chains can even enable the
sampler to converge in situations where it otherwise would not.

We can see this effect in Figure~\ref{fig:scaling_table}. Singling out
the column corresponding to~$D=200$ and~$K=3C$, we notice that using
either~$20$ or~$40$ cores, GESS fails to estimate~$\boldsymbol\sigma$,
indeed the Markov chain fails to burn in during the allotted time (the
average squared errors are~$10^{3.1}$ and~$10^{2.7}$
respectively). However, once we increase the number of cores
to~$60$,~$80$, and~$100$, GESS provides an accurate estimate
of~$\boldsymbol\sigma$ (the average squared errors
are~$10^{-0.6}$,~$10^{-1.7}$, and~$10^{-2.3}$ respectively). In this
case, increasing the number of cores enabled our estimator to
converge. This property contrasts sharply with many other approaches
to parallel sampling. If a single Markov chain running MH will not
converge, then one-hundred chains running MH will not converge either.

%\FloatBarrier
\section{Discussion}
\label{discussion}

In this paper, we generalized elliptical slice sampling to handle
arbitrary continuous distributions using a scale mixture of Gaussians
to approximate the target distribution. We then showed that
parallelism can be used to dynamically choose the parameters of the
scale mixture of Gaussians in a way that encodes information about the
shape of the target distribution in the transition operator. The
result is Markov chain Monte Carlo algorithm with a number of
desirable properties. In particular, it mixes well in the presence of
strong dependence, it does not require hand tuning, and it can be
parallelized over hundreds of cores.

We compared our algorithm to several other parallel MCMC algorithms in
a variety of settings. We found that generalized elliptical slice
sampling (GESS) mixed more rapidly than the other algorithms on a
variety of distributions, and we found evidence that the performance
of GESS can scale superlinearly in the number of available cores.

One possible area of future work is reducing the overhead from the
information sharing. In Section~\ref{reusing_approx} we remarked that
the synchronization requirement leads to faster chains waiting for
slower chains. There are a number of factors which contribute to the
difference in speed from chain to chain. Most obviously, some chains
may be running on faster machines. More subtly, the slice sampling
procedure performs a variable number of function evaluations per
update, and the average number of required updates may be a function
of location. For instance, Markov chains whose current states lie in
narrow portions of the distribution may require more function
evaluations per update. In each situation, the chains with the rapid
updates end up waiting for the chains with the slower updates, leaving
some processors idle. We imagine that a cleverly-engineered system
would be able to account for the potentially different update speeds,
perhaps by sending the chains in the narrower parts of the
distribution to the faster machines or by allowing the slower chains
to spawn multiple threads. Properly done, the performance gain in
wall-clock time due to using GESS should approach the gain as measured
by function evaluations.

In addition to using parallelism to distribute the computational load
of MCMC, we saw that our algorithm was able to use information from
the parallel chains to speed up mixing. One area of future work is
extending the algorithm to take advantage of a greater number of
cores. The magnitude of this performance gain depends on the accuracy
of our multivariate~$t$ approximation, which will increase, to a
point, as the number of available cores grows. However, there is a
limit to how well a multivariate~$t$ distribution can approximate an
arbitrary distribution. We chose to use the multivariate~$t$
distribution because it has the flexibility to capture the general
allocation of probability mass of a given distribution, but it is too
coarse to capture more complex features such as the locations of
multiple modes. After some point, the approximation will not
significantly improve. A more general approach would be to use a
scale-location mixture of Gaussians, which could accurately
approximate a much larger class of distributions. The idea of
approximating the target distribution with a mixture of Gaussians has
been explored by \citet{Schmidler2010} in the context of adaptive
Metropolis--Hastings. We leave it to future work to explore this more
general setting.

% Acknowledgements should go at the end, before appendices and references

\acks{We thank Barbara Engelhardt for kindly sharing the simulated
  genetics data. We thank Eddie Kohler, Michael Gelbart, and Oren
  Rippel for valuable discussions. This work was funded by DARPA Young
  Faculty Award N66001-12-1-4219 and an Amazon AWS in Research grant.}

\FloatBarrier
\appendix
\section*{Appendix A}
\label{sec:appendix}

In Algorithm~\ref{alg:t-params}, we detail the algorithm for estimating the maximum likelihood multivariate~$t$ parameters~$\nu$,~$\boldsymbol\mu$,~$\boldsymbol\Sigma$ from \citet{Liu1995}.

\begin{algorithm}
\caption{Computing the maximum likelihood multivariate~$t$ parameters}
\label{alg:t-params}
\begin{algorithmic}[1]
\Require $I$ points~${\bf x}_i$ (each $D$ dimensional)
\Ensure Maximum likelihood multivariate~$t$ parameters~$\nu$,~$\boldsymbol\mu$,~$\boldsymbol\Sigma$
\State $t \leftarrow 0$
\State Initialize $\nu^{(0)}$, $\boldsymbol\mu^{(0)}$, and $\boldsymbol\Sigma^{(0)}$
\While{$|\nu^{(t)} - \nu^{(t-1)}| < \epsilon$}
  \State Compute the distances from each point ${\bf x}_i$ to $\boldsymbol\mu^{(t)}$ with respect to $\boldsymbol\Sigma^{(t)}$
\begin{equation*}
  \delta_i^{(t)} = \left({\bf x}_i - \boldsymbol\mu^{(t)}\right)^{\mathsf T}\left(\boldsymbol\Sigma^{(t)}\right)^{-1}\left({\bf x}_i - \boldsymbol\mu^{(t)}\right)
\end{equation*}

\State Set
\begin{equation*}
  w_i^{(t+1)} = \frac{\nu^{(t)} + D}{\nu^{(t)} + \delta_i^{(t)}}
\end{equation*}

\State Update the mean and covariance parameters via
\begin{eqnarray*}
  \boldsymbol\mu^{(t+1)} & = &\frac{\sum_{i=1}^I w_i^{(t+1)} {\bf x}_i}{ \sum_{i=1}^I w_i^{(t+1)}} \\
  \boldsymbol\Sigma^{(t+1)} & = & \frac{1}{I}\sum_{i=1}^I w_i^{(t+1)}\left( {\bf x}_i - \boldsymbol\mu^{(t)}\right) \left({\bf x}_i - \boldsymbol\mu^{(t)}\right)^{\mathsf T}
\end{eqnarray*}

\State Using the updated mean and covariance parameters, recompute the distance
\begin{equation*}
  \delta_i^{(t+1)} = \left({\bf x}_i - \boldsymbol\mu^{(t+1)}\right)^{\mathsf T}\left(\boldsymbol\Sigma^{(t+1)}\right)^{-1}\left({\bf x}_i - \boldsymbol\mu^{(t+1)}\right)
\end{equation*}

\State Let $\psi$ be the digamma function, and let \begin{equation*}
  w_i = \frac{\nu + D}{\nu + \delta_i^{(t+1)}}
\end{equation*}

\State Set $\nu^{(t+1)}$ by solving for $\nu$ in the equation
\begin{equation*}
-\psi\left(\frac{\nu}{2}\right) + \log\left(\frac{\nu}{2}\right) + \frac{1}{I} \sum_{i=1}^I \left(\log\left(w_i\right) - w_i\right) +  \psi\left( \frac{\nu + D}{2} \right) - \log\left( \frac{\nu + D}{2} \right) = -1
\end{equation*}

\State $t \leftarrow t + 1$

\EndWhile

\State \Return $\nu^{(t)}$, $\boldsymbol\mu^{(t)}$, and $\boldsymbol\Sigma^{(t)}$

\end{algorithmic}
\end{algorithm}

\FloatBarrier
\vskip 0.2in
\bibliography{refs}

\end{document}